\documentclass[11pt,twoside]{article}

\usepackage{asp2006}
\usepackage{epsf}
\usepackage{graphicx}
\usepackage{lscape}

\begin{document}
\title{Sun-bathing around low-mass protostars: APEX-CHAMP$^+$
observations of high-$J$ CO} 

\author{Ewine F.\ van
Dishoeck$^{1,2}$, Tim A.\ van Kempen$^{1,3}$ and Rolf G\"usten$^{4}$} 

\affil{$^1$Leiden Observatory, Leiden University, P.O. Box 9513, 2300
RA Leiden, The Netherlands; ewine@strw.leidenuniv.nl}
\affil{$^2$Max-Planck Institut f\"ur Extraterrestrische Physik (MPE),
Giessenbachstrasse 1, 85748 Garching, Germany}
\affil{$^3$Harvard-Smithsonian Center for Astrophysics, Cambridge, MA 02138, USA}
%
\affil{
$^4$Max Planck Institut f\"ur Radioastronomie, Auf dem H\"ugel 69, D-53121, Bonn, Germany}

\begin{abstract} 
  We present the first maps of high-excitation CO $J$=6--5 and 7--6
  and isotopologue lines over $\sim 2'-5'$ regions at 10$''$
  resolution toward low-mass protostars to probe the origin of the
  warm gas in their surroundings. The data were obtained using the
  CHAMP$^+$ 650/850 GHz heterodyne array receiver on
  APEX.\footnote[5]{This publication is based on data acquired with
    the Atacama Pathfinder Experiment (APEX). APEX is a collaboration
    between the Max-Planck-Institut f\"ur Radioastronomie, the
    European Southern Observatory, and the Onsala Space Observatory.}
  Surprisingly strong {\it quiescent extended} narrow-line high-$J$
  $^{12}$CO 6--5 and 7--6 emission is seen toward all protostars,
  suggesting that heating by UV photons along the outflow cavity
  dominates the emission.  At the source position itself, passive
  heating of the collapsing inner envelope by the luminosity of the
  source also contributes. The UV photons are generally not energetic
  enough to dissociate CO since the [C I] 2--1 emission, also probed
  by our data, is weak except at the bow shock at the tip of the
  outflow. The extended UV radiation is produced by the star-disk
  boundary layer as well as the jet- and bow-shocks, and will also
  affect the chemistry of species such as H$_2$O and HCN around the
  outflow axis. Shock-heated warm gas characterized by broad CO line
  profiles is seen only toward the more massive Class 0
  outflows. Outflow temperatures, estimated from the CO 6--5/3--2 line
  wing ratios, are $\sim$100 K. These data illustrate the importance
  of getting spatial information to characterize the physical
  processes in YSO surroundings. Such information will be important to
  interpret future {\it Herschel} and ALMA data.

\end{abstract}



\section{Introduction}

During the formation of low-mass stars, material moves from the cloud
core to the collapsing envelope and disk onto the growing star. At the
same time, part of the envelope is dispersed by the jets and winds
from the protostar, limiting its growth.  Characterizing these
different physical components of low-mass young stellar objects (YSOs)
(envelope, disk, outflow, cloud) is thus crucial for understanding the
physical evolution of the earliest stages of star formation: this is
the phase in which the final mass of the star and disk are determined
and the conditions for planet formation are set.  Because of the high
extinction ($A_V>50$ mag), these sources are best studied at
(sub)millimeter and far-infrared wavelengths. Most studies to date
have focussed on the cold gas and dust around protostars, in
particular using the dust continuum and low-$J$ lines of CO and its
isotopologues $^{13}$CO, C$^{18}$O and/or C$^{17}$O
\citep{Shirley00, Jorgensen02, Jorgensen05a, Hatchell09}.  In contrast,
the warm gas is much more diagnostic of the energetic processes that
shape these deeply embedded sources.

The most direct probes of warm (50--200 K) gas are the high-$J$
($J\geq 4$) lines of CO (for example, $E_u=115$~K for $J$=6).  Complex
molecules such as H$_2$CO and CH$_3$OH also have high excitation lines
in the 230 and 345 GHz atmospheric windows which have been detected
toward low-mass protostars \citep[e.g.][]{vanDishoeck95,Blake95,
  Ceccarelli00,Schoeier02,
  Maret04,Jorgensen04,Bottinelli04,Bottinelli07}, but their strong
abundance variations complicate their use as tracers of the physical
structure.  Observations of high-excitation lines of CO can be carried
out from the ground in the atmospheric windows at 650 and 850 GHz, but
they require excellent weather conditions.  Such pioneering data were
obtained more than a decade ago at single positions with the Caltech
Submillimeter Observatory (CSO) using its excellent receiver suite
\citep{Hogerheijde98} and with the James Clerk Maxwell Telescope
(JCMT) \citep{Schuster93,Schuster95}.  Around the same time, the {\it
  Infrared Space Observatory} (ISO) provided spectrally unresolved
data on even higher-$J$ lines in a large $\sim 80''$ beam
\citep[e.g.,][]{Ceccarelli98,Giannini99, Nisini99,Giannini01,
  Nisini02}.  The lack of spatial and/or spectral information on the
warm gas prevented an in-depth analysis of these data, however.

As a result, the location and heating mechanisms of warm dense gas
near low-mass protostars are still strongly debated
\citep[e.g.,][]{Nisini00,Ceccarelli99,Maret02}.
Proposed options include (i) passive heating of the inner $\sim$100 AU
region of the collapsing envelope by the protostellar luminosity; (ii)
active heating in shocks created by the interaction of jets and winds
from the protostar with the envelope and cloud out to large distances;
(iii) heating by UV photons escaping through outflow cavities and
scattered back into the envelope on a few thousand AU scales
\citep{Spaans95};
and (iv) a forming protoplanetary disk heated by accretion shocks
\citep{Ceccarelli02}.
In scenarios (i) and (iv), the high-$J$ CO emission should be
spatially unresolved and centered on the source. In scenario (ii),
extended broad emission lines are expected to be seen out to the tip
of the outflow. In scenario (iii), spatially extended narrow high-$J$
lines are expected.  APEX-CHAMP$^+$, with its heterodyne spectral
resolution, 7--10$''$ beam ($\sim 1000$~AU at 120 pc) and mapping
capabilities, can directly test the various scenarios. Located at the
Chajnantor plateau in Northern Chile with excellent atmospheric
transparency, it allows much more routine observations of these
high-$J$ lines than possible from Mauna Kea.  We report here the
initial results of a survey of high$-J$ CO lines toward a set of
low-mass protostars (van Kempen et al.\ 2009a,b).


\section{Observations}

The CHAMP$^+$ instrument, developed jointly by the MPIfR and SRON
Groningen, is the only submillimeter array receiver in the world able
to simultaneously observe molecular line emission in the 650 and 850
GHz atmospheric windows on arcminute spatial scales with 7--9$''$
pixels \citep{Kasemann06,Guesten08}.  It is the successor of the CHAMP
array at 460 GHz deployed on the CSO until 2003.  CHAMP$^+$ has 14
pixels (7 in each frequency window) arranged in a hexagon of 6 pixels
around 1 central pixel.  To obtain fully sampled maps, the array was
either moved in a small hexagonal pattern, or, more efficiently, in an
on-the-fly (OTF) mode. The hexa-pattern covers a region of about
30$''\times30''$, whereas small OTF maps are 40$''\times40''$, or
occasionally larger.

The observations were obtained during several observing runs between
June 2007 and November 2008 in three different line settings, with the
combination of lines chosen to match required integration times: (1)
$^{12}$CO $J$=6--5 and 7--6; (2) $^{13}$CO 6--5 and [C I] 2--1; and
(3) C$^{18}$O 6-5 and $^{13}$CO 8--7. For the latter setting (so far
done only for HH 46), a stare mode was used to increase the $S/N$ on
the central pixel.  A position switch of 900$''$ or larger was used
for all settings, except for the stare setting (3), which used a
beam-switching of 90$''$. Complementary lower-$J$ lines of CO, HCO$^+$
and their isotopologues have been obtained with the facility APEX-1
and APEX-2a receivers.

During the observations in 2007, the backend consisted of two Fast
Fourier Transform Spectrometer (FFTS) units serving the central pixel
(resolution up to 0.12 MHz), and 12 MPI-Auto-Correlator Spectrometer
(MACS) units (resolution 1 MHz) connected to the other pixels.  Since
July 2008, all pixels are connected to FFTS units.  Main beam
efficiencies, derived using observations on planets, are 0.56 at 650
GHz and 0.43 for 850 GHz. Typical single sideband system temperatures
are 700 K and 2100 K, respectively.  Pointing was checked on various
planets and sources and was found to be within 3$''$. Calibration is
estimated to have an uncertainty of 30$\%$ for both frequencies.
Integration times were such that the typical rms in a 0.7 km s$^{-1}$
velocity bin at 650 GHz in setting (1) is 0.2-0.4 K, in setting (2)
0.2 K, and setting (3) 0.1 K, respectively. At the map edges, noise
levels are often higher due to the shape of the CHAMP$^+$ array.

For comparison, the {\it Herschel Space Observatory} will allow
observations of far-IR CO lines with the PACS instrument at spatial
resolutions similar to these APEX data ($\sim$10$''$).  At its longer
wavelengths ($\sim$500 $\mu$m), the beam of {\it Herschel} is
comparable to, or smaller than, the field of view of CHAMP$^+$ ($\sim
40''$). Thus the CHAMP$^+$ data provide important complementary
information on the distribution of warm gas within the {\it
  Herschel-HIFI} beams.

\begin{figure}[t]
\begin{center}
\includegraphics[width=12cm]{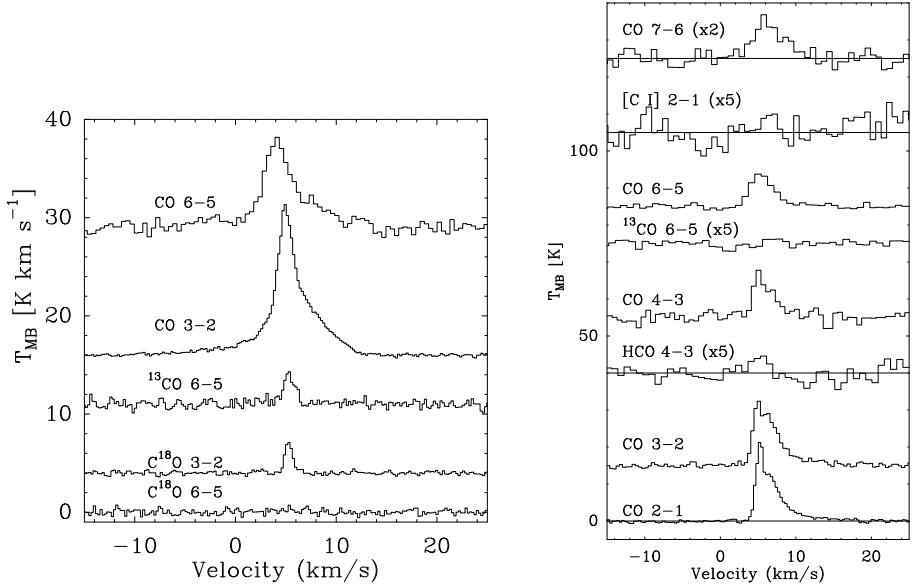}
\caption{Left: Spectra of various CO and isotopologue lines centered
  at the source position. Right: Spectra at an off position in the red
  outflow lobe at $(-20'',-20'')$.  The quiescent gas has a FWHM of
  1.5 km s$^{-1}$ centered at $V_{\rm LSR}$=5.3 km s$^{-1}$ (based on
  van Kempen et al.\ 2009a, published with permission). }
\label{fig:hh46-spec}
\end{center}
\end{figure}

\begin{table}
\caption{Sample of sources observed with CHAMP$^+$ as of summer 2009}
\label{tab:sources}
\small
\begin{center}
\begin{tabular}{l l l l l r r r r}
\hline \hline 
\noalign{\smallskip}
Source & \ \ RA & \ \ Dec & \ $D$ & $L_{\rm{bol}}$ & $T_{\rm{bol}}$ 
& Class & Ref.   \\
 & (J2000) & (J2000) & (pc)& ($L_{\rm{\odot}}$) & (K) &   \\ \hline 
\noalign{\smallskip}
NGC 1333 IRAS2 & 03:28:55.2& +31:14:35  & 250 & 12.7 & 62  & 0 & 2 \\ 
NGC 1333 IRAS4A/B & 03:29:11& +31:13  & 250 &8 & 35  & 0 & 3 \\ 
L1551 IRS5     & 04:31:34.1&+18:08:05.0 & 160 & 20   & 75  & 1 & 2\\
TMR 1           & 04:39:13.7& +25:53:21  & 140 & 3.1 & 133  & 1 & 2 \\  
HH 46           & 08:25:43.8& -51:00:35.6& 450 & 16 & 102   & 1 & 1 \\
Ced 110 IRS4   & 11:06:47.0& -77:22:32.4& 130 & 0.8 & 55   & 1 & 2 \\  
BHR 71          & 12:01:36.3& -65:08:44  & 200 & 11 & 60    & 0 & 2,5\\ 
IRAS 12496-7650 & 12:53:17.2& -77:07:10.6& 250 & 24 &325 & 1 & 2 \\
Serpens core &  18:29:55 & +01:14 & 250 & 30/5 & - & 0 & 4 \\    
RCrA IRS7 & 19:01:55 & -36:57:21 & 170 & - & -   & 0 & 2  \\ 
\hline
\end{tabular}
{\smallskip}
References: (1) van Kempen et al.\ (2009a); (2) van Kempen et al.\ (2009b);
(3) Yildiz et al.\ in prep.; (4) and (5) Kristensen et al.\ in prep.
\end{center}
\end{table}

The sample studied so far consists of a dozen well-known and
well-studied embedded protostars that can be observed from the
southern sky.  All sources have been studied in previous surveys of
embedded YSOs \citep[e.g.,][]{Jorgensen02,Jorgensen04,Groppi07}.
Table~\ref{tab:sources} gives the parameters of each source and its
properties. Initial CHAMP$^+$ results have been published by
\citet{Kempen09a} for HH 46 and \citet{Kempen09b} for a larger
sample. The data on NGC 1333 IRAS 4A/B (Yildiz et al., in prep.)
and the Serpens core containing SMM1, SMM3 and SMM4 (Kristensen et
al., in prep.) are being analyzed. Results using the single-pixel
460/850 GHz FLASH receiver on APEX can be found in
\citet{Kempen06,Kempen09c} (see also van Kempen 2008).

\section{HH 46 as an example}

\subsection{The source}

The first source targetted with APEX-CHAMP$^+$ was HH 46, located at the
edge of an isolated Bok globule ($D=$ 450 pc) \citep{Schwartz77}. It
is well-known for its spectacular outflow, observed at both visible
and infrared wavelengths with the {\it Hubble} and {\it Spitzer
Space Telescope}
\citep[e.g.,][]{Heathcote96,Stanke99,Noriega04,Velusamy07}. Deep
H$\alpha$ observations have revealed bow shocks associated with the HH
46 outflow up to a parsec away from the central source
\citep{Stanke99}.  Its blue-shifted lobe expands into a low density
region outside the cloud, whereas the red-shifted lobe plows into the
dense core.  The internal driving source for the flow is HH~46 IRS1
($L=16$ $L_{\rm\odot}$) \citep{Raymond94,Schwartz03}.
\citet{Chernin91} and \citet{Olberg92} mapped this region using low
excitation CO lines showing that the molecular emission in the red-shifted
outflow lobe is indeed much stronger than that in the blue-shifted
one.

\begin{figure}[t]
\begin{center}
\includegraphics[width=12cm]{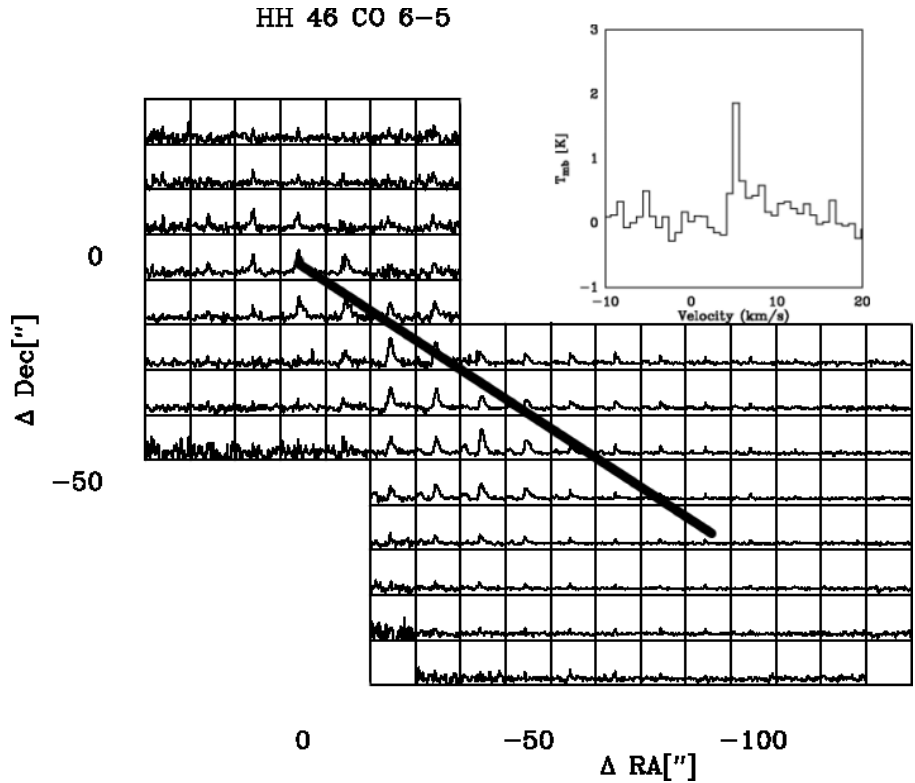}
\caption{Spectral map of the $^{12}$CO J=6--5 lines obtained with
  APEX-CHAMP$^+$. The thick line indicates the outflow direction. Note
  the narrow CO line profiles offset from the outflow walls, in
  addition to the broad outflowing gas (see blow-up of spectrum at
  $-50'',-60''$)  (based on van Kempen et al.\ 2009a). }
\label{fig:hh46-specmap}
\end{center}
\end{figure}

\subsection{Results}

Figure~\ref{fig:hh46-spec} (left) shows the high-quality spectra taken at
the position of HH 46 IRS1. Emission is clearly detected for all
lines, except for C$^{18}$O $J$=6--5 and $^{13}$CO 8-7. [C I] 2--1
emission is surprisingly weak at the central source, only $T_{\rm
MB}=2.7$ K. Line wings are clearly seen for the $^{12}$CO lines but
become less prominent relative to the narrow line component for the
higher excitation lines. The same is seen at outflow positions away from
the source (Fig.~\ref{fig:hh46-spec}, right)

Figure~\ref{fig:hh46-specmap} presents the $^{12}$CO 6-5 spectra
binned to square $10''\times 10''$ pixels.  In addition to the line
wings characterizing the swept-up outflow gas, surprisingly strong
{\it extended narrow} CO lines are seen along the direction of the
outflow axis (see for example spectra at $-30'',10''$ and
$-30'',-30''$).  These data directly support scenario (iii), in which
UV radiation heats the surrounding outflow walls.

\begin{figure}[t]
\begin{center}
\includegraphics[scale=0.4,angle=-90]{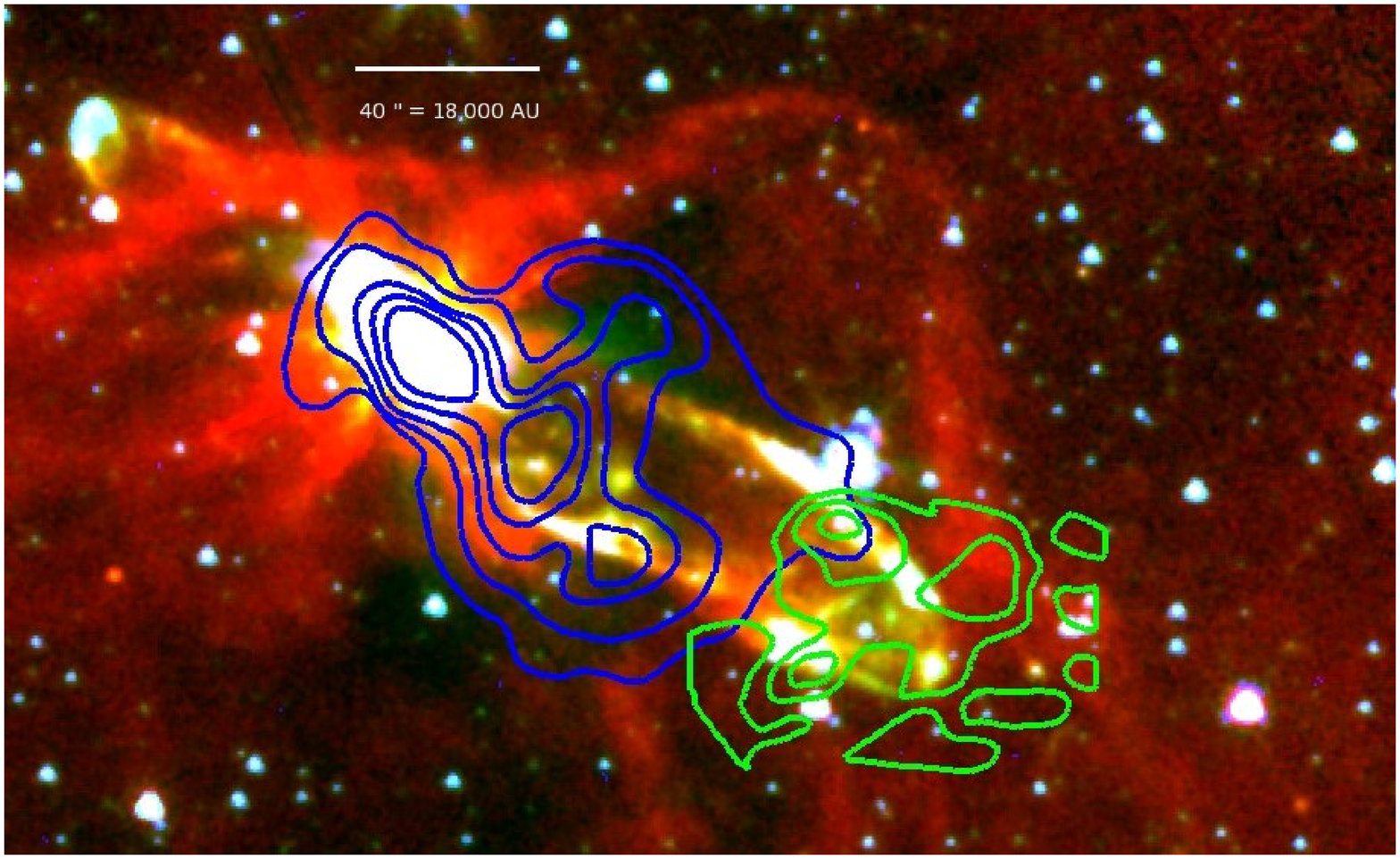}
\caption{{\it Spitzer} three color (3.6 (blue), 4.5 (green) and 8 $\mu$m
  (red)) image from \citet{Noriega04} with the contours of integrated
  CO $J$=6--5 emission (blue/dark) observed with APEX-CHAMP$^+$
  overlaid. The CO contours are in steps of 5 K km s$^{-1}$. The [C I]
  2--1 emission (green/light) is detected weakly on source but peaks
  further down at the bow shock at the tip of the outflow where the UV
  photons produced in the fast bow shock are hard enough to dissociate
  CO. Analysis of the line profiles shows that the emission consists
  both of accelerated swept-up gas along the outflow as well as
  quiescent, photon-heated gas surrounding the outflow cavity walls
  (see also Fig.\ 2) (reproduced with permission from van Kempen et
  al.\ 2009a).}
\label{fig:hh46-Spitzer}
\end{center}
\end{figure}

To test this scenario further, the outflow region was mapped more
extensively in $^{12}$CO 6--5 and [C I] 2--1 lines out to the tip of
the outflow characterized by the bow shock around
($-100'',-60''$). Figure~\ref{fig:hh46-Spitzer} shows an overlay of
the CO and [C I] integrated intensity maps on the {\it Spitzer}
image. The CO emission (both narrow and broad) clearly disappears
toward the location of the bow shock.  Although the dynamic range of
the [C I] map is not large, narrow [C I] 2--1 line emission is
somewhat enhanced around the bow shock position. This suggests that
along most of the outflow axis the UV photons are energetic enough
($>$few eV) to heat the gas, but not to photodissociate CO. In
contrast, at the bow-shock position, photons must be present with high
enough energies to dissociate CO ($>$11 eV, $<$1100 \AA) and produce
[C I] emission both ahead and just behind the shock. According to the
models by \citet{Neufeld89}, $J-$shocks with velocities larger than 90
km s$^{-1}$ produce CO dissociating photons, mostly through the
two-photon decay of metastable He(2 $^1$S) atoms. At the bow shock,
the shock velocity is measured to be 220 km s$^{-1}$ from atomic lines
\citep{Fernandes00}, more than sufficient to produce these high energy
photons. Because of the high densities in the core, they can travel
only a short path before being absorbed by dust, H$_2$ or CO. This
situation is reminescent of the IC 443 supernova remnant, where
quiescent [C I] emission was also detected ahead of the shock and
interpreted to arise from photodissociation of CO in the pre-shocked
gas \citep{Keene96}.

\begin{figure}[t]
\begin{center}
\includegraphics[scale=0.28]{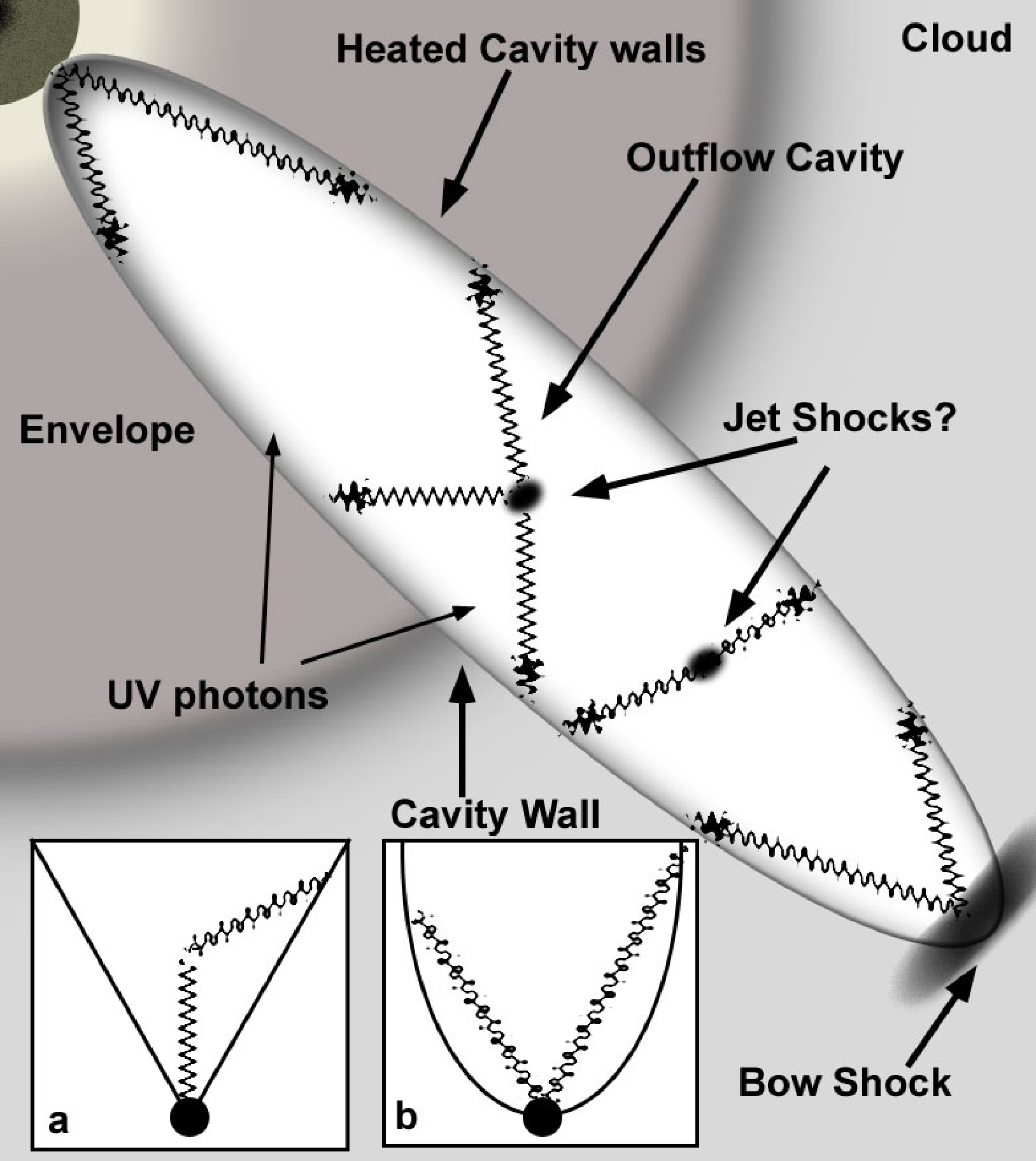}
\caption{Cartoon of the HH 46 outflow on scales up to the
  bow shock at $\sim$60,000 AU from the source, illustrating the
  photon heating of the cavity walls (either directly or scattered by
  dust) responsible for the bulk of the quiescent high-$J$ CO
  emission. UV photons are created at the star-disk accretion boundary
  layer as well as in the bow- and jet shocks.  The inserts illustrate
  possible geometries at the base of the outflow, with the parabolic
  shape allowing more direct illumination of the outflow walls
  (reproduced with permission from van Kempen et al. 2009a). }
\label{fig:hh46-cartoon}
\end{center}
\end{figure}

The CHAMP$^+$ data thus provide support for scenario (iii), but with
an extension of the original model by \citet{Spaans95}. First, the UV
photons originate not only from the source itself but are also
produced further out by the bow- and jet-shocks. Second,
\citet{Spaans95} suggested that the UV photons from the star-disk
accretion boundary layer reach large distances by scattering on dust
grains in the outflow cone. \citet{Bruderer09} and \citet{Kempen09a}
note that direct illumination of the outflow walls may be more
effective if the outflow walls have a parabolic rather than a linear
straight shape (see Figure~\ref{fig:hh46-cartoon} for illustration).

To quantify the contribution to the emission at the source position
from scenario (i), a passively heated envelope model has been
constructed following the method of \citet{Jorgensen02}. The envelope
is assumed to have a power-law density structure, with the temperature
determined by a dust radiative transfer calculation with the
luminosity of the source as input. The power-law exponent, mass and
extent of the envelope are determined by a best fit to the spectral
energy distribution and dust continuum maps. For HH 46, a LABOCA 850
$\mu$m map over the central $5'\times 5'$ map has been obtained for
this purpose. Subsequently, the CO excitation and line emission are
calculated for the resulting best-fit temperature and density envelope
structure using the Monte Carlo radiative transfer package of 
\citet{Hogerheijde00}, assuming that the gas temperature equals the dust
temperature. The latter assumption should hold for the inner envelope
where densities are high enough ($>10^6$ cm$^{-3}$) for gas and dust
to be thermally coupled.

This quantitative analysis shows that the envelope emission can
reproduce some of the observed high-$J$ CO emission shown
Figure~\ref{fig:hh46-spec} (left), but generally underproduces the
$^{12}$CO and $^{13}$CO 6--5 lines by nearly a factor of 3.
Increasing the CO abundance is not possible since then the C$^{18}$O
3--2 and 6--5 lines would be overproduced. Thus, an additional
optically thin but warm component needs to be added to the model to
explain the data. It needs to be relatively unobscured since it adds
emission to that of the optically thick $^{12}$CO lines from the inner
envelope. Photon-heated gas in the (outer) envelope fulfills these
criteria.  The weak observed [C I] emission at the source position can
be accounted for by a low atomic C abundance ($\sim$few $\times
10^{-7}$ with respect to hydrogen) that is maintained by cosmic ray
induced UV photodissocation of CO inside the dense envelope.

The current data do not provide constraints on scenario (iv), i.e.,
whether some of the hot gas could be due to the accretion shock onto
the disk in the embedded phase. To answer this question, submillimeter
interferometry in the CO $J$=6--5 and 7--6 lines with ALMA is
needed. High spatial and spectral resolution mid-infrared CO $v$=1--0
data at 4.6 $\mu$m can also provide insight into this question.

In summary, the quantitative analysis by \citet{Kempen09a} shows that
the warm gas traced by the high-$J$ CO arises from three different
components: (i) the quiescent inner warm envelope, accounting for
roughly 30\% of the $^{12}$CO and $^{13}$CO on-source intensities;
(ii) the outflowing gas, accounting for the extended red and blue
wings; and (iii) {\it UV-heated quiescent gas ($T\geq 100$ K),
  accounting (surprisingly!)  for the bulk of the extended narrow 6--5
  and 7--6 emission.}


\section{Other sources}

Data for several of the other sources in Table~\ref{tab:sources} are
presented in \citet{Kempen09b} and confirm the results found for HH
46.  Warm gas, as traced by $^{12}$CO 6--5 and 7--6, is present in all
observed protostars at the central position.  For Ced 110 IRS 4, a
quantitative model similar to that for HH 46 shows that passive
heating of the envelope is again insufficient to explain the observed
$^{12}$CO 6--5, 7--6 and $^{13}$CO 6--5 lines, requiring heating of
the envelope by UV photons even at the (0,0) position.

\begin{figure}[t]
\begin{center}
\includegraphics[scale=0.5,angle=-90]{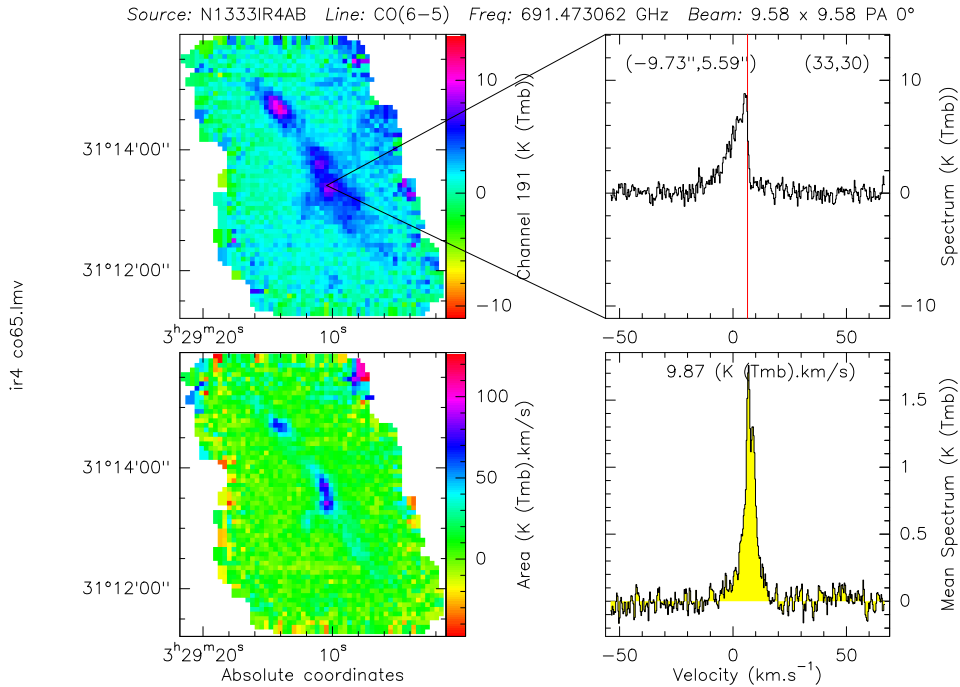}
\caption{Preliminary CO $J$=6--5 map of the NGC 1333 IRAS 4A/B region.
  The top map shows a channel image at the central velocity whereas
  the lower map shows an integrated intensity map. Two representative
  spectra are shown as well (Yildiz et al., in prep.).}
\label{fig:ngc1333-map}
\end{center}
\end{figure}

Mapping shows that photon heating of the cavity walls takes place on
arcmin scales for the outflows of several other Class 0 and Class I
protostars, as seen at positions off-source of BHR 71, NGC 1333 IRS2,
and L~1551 IRS5. Figure~\ref{fig:ngc1333-map} shows a preliminary map
of the NGC 1333 IRAS4A/B region, with both outflow and narrow emission
present along the outflow axis of the IRAS4A flow (Yildiz et al., in
prep.).  As for HH 46, the necessary UV photons are likely created by
internal jet shocks and the bow shock where the jet interacts with the
ambient medium, in addition to the disk-star accretion boundary
layer. The distribution of the quiescent CO 6--5 and 7--6 emission
seen toward BHR~71, where the narrow emission is stronger at larger
distances from the source, confirms the hypothesis that UV photons
necessary for heating can originate from both mechanisms. Although
difficult to disentangle in cases where the outflow is not in the
plane of the sky, the observations suggest that photon heating is a
common phenomena along all outflows.  The lack of [C I] 2--1 emission
in the outflows indicates that there is limited production of
energetic CO dissociating photons in the shocks.

Shocked broad $^{12}$CO 6--5 and 7--6 lines are only seen prominently
in the flows of more massive sources (NGC 1333 IRAS2 and IRAS4A/B, BHR
71, HH~46, RCrA IRS7 and the Serpens sources), while lower mass flows
do not show significant $^{12}$CO 6--5 and 7--6 line wings.
\citet{Bachiller99} and \citet{Arce06} show that the outflows of Class
I sources are more evolved and are driven with much less energy,
producing significantly weaker shocks and less swept-up gas \citep[see
also][]{Bontemps96,Hogerheijde98}.  This is also reflected in the
decreasing maximum velocity of $^{12}$CO 6--5 with lower outflow
force.  From the CO 6--5/3--2 line ratios, kinetic temperatures of
$\sim$100 K are found for the molecular gas in the flows studied
here. Such temperatures agree with expected conditions of bow-shock
driven shell models by \citet{Hatchell99}, in which the
momentum-conserving shells expand and loose kinetic energy to heat the
swept-up molecular gas. Only the flows of L~1551 IRS5 and IRAS
12496-7650 appear to be colder ($<$50 K), consistent with the
`fossil', empty nature of these flows.

The intensities at all positions toward RCrA IRS7 are an order of
magnitude higher than can be produced by passive heating, and probably
result from a significant PDR region near the source. Likely, RCrA
itself heats the outside of the RCrA IRS7 envelope and cloud region.

\section{Concluding remarks}

Maps in high-$J$ CO submillimeter lines clearly provide crucial
information to determine the location and origin of warm dense gas
near YSOs and the feedback that the young star has on its
surroundings. Questions to be addressed with future larger samples
combined with {\it Herschel} data include: is the UV-heated, quiescent
component always dominant, or does this change from the deeply
embedded Class 0 phase to the less embedded Class 1 phase and
eventually the disk-dominated phase? Does this heating limit the
ability of the cloud to collapse further and thus help determine the
final mass of the star? How hot is the outflowing molecular gas and
how does this depend on evolutionary state? 

The processes identified here have strong implications not only for
the physical evolution, but also for the chemistry near protostars,
especially the origin of the emission from water and complex organic
molecules. For example, if UV radiation is indeed widespread around
low-mass YSOs, it will readily dissociate H$_2$O into OH and H,
strongly affecting the predictions and interpretation of upcoming {\it
  Herschel-HIFI} data.  Some indications of this process have already
been found through enhanced narrow CN emission (a photoproduct of HCN)
along the outflow axis of the low-mass YSO L483
\citep{Jorgensen04}. In addition, the UV photons can photoprocess and
photodesorb ices such as CH$_3$OH efficiently and thus enhance their
gas phase abundances (\"Oberg et al.\ 2009a,b).  Recent mapping of the
CHAMP$^+$ sources in HCN, CN and CH$_3$OH with the JCMT-HARP and IRAM
30-m HERA arrays is in progress to test these chemical consequences
(Kristensen et al., in prep).

Together, the quality and richness of these data sets are a tribute to
the instrument builders and illustrate how far and rapidly the field
has developed since the first pioneering 230 GHz data by
\citet{Phillips73}.

\acknowledgements None of these results would have been possible
without the drive of Tom Phillips to develop the field of
submillimeter astronomy and its instrumentation. The authors enjoyed
many lively discussions with Tom over the past decades, always
stimulating them to obtain the best possible data and driving them to
seek a deeper physical and chemical understanding of the results.

They are also grateful to A. Baryshev, A.\ Belloche, W. Boland,
M. Hogerheijde, L.\ Kristensen, K.\ Menten, P. Schilke, R.\ Stark, F.\
Wyrowski and U.\ Yildiz for building the CHAMP$^+$ instrument and/or
assisting with the observations and analysis. The Dutch contribution
to CHAMP$^+$ was financed by the Netherlands Organization for Scientific
Research (NWO) under grant 614.041.004 (PI: W. Boland), and managed by
the Netherlands Research School for Astronomy (NOVA).



\end{document}